# THREE'S A CROWD: ON CAUSES, ENTROPY AND PHYSICAL ESCHATOLOGY

Milan M. Ćirković

*Astronomical Observatory Belgrade*

*Volgina 7, 11160 Belgrade, SERBIA AND MONTENEGRO*

*e-mail:* `arioch@eunet.yu`

Vesna Milošević-Zdjelar

*Department of Physics and Astronomy*

*University of Manitoba, Winnipeg MB,  R3T 2N2, CANADA*

*e-mail:* `ummilose@cc.UManitoba.CA`

# THREE'S A CROWD: ON CAUSES, ENTROPY AND PHYSICAL ESCHATOLOGY

**Abstract.** Recent discussions of the origins of the thermodynamical asymmetry ("the arrow of time") by Huw Price and others are critically assessed. This serves as a motivation for consideration of relationship between thermodynamical and cosmological causes. Although the project of clarification of the thermodynamical explanandum is certainly welcome, Price excludes another interesting option, at least as viable as the sort of Acausal-Particular approach he favors, and arguably more in the spirit of Boltzmann himself. Thus, the competition of explanatory projects includes three horses, not two. In addition, it is the Acausal-Particular approach that could benefit enormously from dissociation from fanciful ideas of low-entropy future boundary conditions entertained by Price. Novel revolutionary developments in observational cosmology, as well as in the nascent astrophysical discipline of physical eschatology, have obliterated such hypotheses. Also, the Acausal-Anthropic approach we propose, offers another clear instance of disteleological nature of the anthropic principle.

## 1. Introduction

Wherever and whenever two well-trenched alternatives exist (be it the Cold War, the great cosmological debate of 1950s, contemporary "Science Wars," or affirmative action), dissenting voices suggesting the third way—an option not in conventional bipolar taxonomies—are bound to be easily dismissed. Still, those options certainly should be heard, especially when they are in fact quite old, predating some of the main contenders. This applies to the debate of the nature of explanandum in discussions surrounding the thermodynamical temporal asymmetry ("the thermodynamical arrow"). In recent remarkably clear and interesting papers (Price 2002a,b), Huw Price attempted to show that there are two competing projects for explanation of the perceived thermodynamical asymmetry, which he labeled Causal-General and Acausal-Particular approaches. Furthermore, his intention was to show the superiority of the Acausal-Particular approach, which was in accordance with other pieces of his atemporal worldview presented, for instance, in his recent brilliant monograph on the subject of temporal asymmetries (Price 1996). Closely related (though less clearly exposed and with somewhat different emphasis) views may be found in Zeh (1992), Lieb and Yngvason (1999), and Uffink and Brown (2001).

The problem of the time-asymmetry of thermodynamics is the following. In our experience, systems increase in entropy in the forward direction of time. The underlying dynamical laws which are taken to govern thermodynamic systems—be it Newtonian or relativistic or quantum—however, are symmetric in time: statistical mechanics predicts that entropy is overwhelmingly likely to increase in both temporal directions. So where does the asymmetry of thermodynamics (and of our experience generally) come from? Although variations on the same topic existed since antiquity (e.g., Kutrovátz 2001), the problem in its modern form was probably first formulated by Irish physicist Edward P. Culverwell, who concluded in 1890 that the kinetic theory alone could never succeed in explaining the second law of thermodynamics (Culverwell 1890). As Boltzmann (1964) asked: "is the apparent irreversibility of all known natural processes consistent with the idea that all natural events are possible without restriction?"

The answer Price advocates relies on a low-entropy initial boundary condition: if the initial state of the universe is the one of extremely low entropy, then Boltzmannian statistical considerations yield overwhelmingly likely entropy increase towards the future throughout the history of the universe. Price contrasts this account, a version of what he terms the *Acausal-Particular* approach, with theories which derive the thermodynamic asymmetry from some underlying asymmetric causal or dynamical mechanism (like the quantum-mechanical state reduction), what he calls *Causal-General* views. To Price's mind, these two kinds of account are the only serious contenders for explanations of the time-asymmetry of thermodynamics.

Hereby, we would like to put the problem into a wider perspective. Notably, we demonstrate that Price's account is biased and lacking several important ingredients. This circumstance offers a neat motivation for highlighting important issues in application of the conventional thermodynamical apparatus to cosmological and eschatological context. While the desire for clarification of the common task is highly commendable, it is important that taxonomy is maximally comprehensive *and* just. Price's taxonomy fails in one important respect: it fails to notice an alternative to both Causal-General and Acausal-Particular view. Therefore, in this note, we would like to point out that from the very preferred Price's Acausal-Particular approach, another option branches, which deserves a separate mention in reviewing the ways toward the explanation of the thermodynamical asymmetry. This third approach differs markedly from the other two in its conception of what needs to be done to solve the puzzle. In proposing this, we follow the lead of Price himself who, introducing his two proposals, points out (Price 2002b, sec. 1.1):

> So far as I know, the distinction between these two approaches has not been drawn explicitly by other writers. Without it, it is not easy to appreciate the possibility that many familiar attempts to explain the time-asymmetry of thermodynamics might be not *mistaken* so much as *misconceived*—addressed to the wrong problem, in looking for time-asymmetry in the wrong place.

We shall see that, unfortunately, the same razor cuts both ways, and that there are several instances in which Price's own favorite proposal is misconceived, for *exactly the same reason*: searching for the solution of the puzzle in a wrong place.

Instead, we propose a third way out of the thermodynamical arrow of time quandary, one which relies on the universal observational selection effect, brought about by our existence as intelligent observers. It is an objective proposal for explanation of particular low-entropy initial conditions observed in our cosmological domain, which relies on the observational selection effects, embodied in the so-called anthropic principle (for exhaustive reviews, see Barrow and Tipler 1986; Balashov 1991; Hogan 2000; Bostrom 2002). Anthropic principle(s) have been subject to much controversy in recent years, centered mainly on the alleged teleological implications of at least some of their formulations (e.g. Klee 2002). In our view, these criticisms are *non sequitur*, since the main thrust of the anthropic reasoning is not only disteleological, but indeed anti-teleological. The apparent fine-tuning of our cosmological domain is a puzzling empirical fact, in need of an explanation. The design hypothesis provides one possible (teleological) explanation, though hardly a physical or entirely scientific one, which one may or may not find plausible on other grounds. But the anthropic reasoning, together with other advances in physics, provides an alternative explanation: if there is an ensemble of universes (also called the *multiverse*), with a suitably varying range of properties (and current quantum cosmological models, like Linde's chaotic inflation, do suggest that that is the case), then without any fine-tuning, one would expect some of these universes to be just right for life. Even if such universes are in a small minority, because of an observation selection effect (the life-less universes, where the constants and parameters are "wrong" contain no observers and cannot be observed), such a theory can predict that we should observe precisely what we do in fact observe: a universe that *appears* to be fine-tuned. This will not happen because the underlying theory is fine-tuned (which would be a hallmark of the design hypothesis), but because an observation selection effect guarantees that we observe a very atypical part of the whole of physical existence. The theory of observation selection has recently been given a detailed formulation by Bostrom (2003), and we wholeheartedly direct the interested reader to this monograph for a remarkably clear and coherent treatment of all the issues involved. In the same time, the arguments of Bostrom and others represent the best defense of the epistemological validity of the anthropic thinking, and we shall adopt them in the following discourse (see also the companion paper to the present one, Ćirković 2003).

Finally, we would like to bring some attention to the fact which Price does his best to avoid: that even if one accepts the Acausal-Particular approach on which it is true that entropy does not *necessarily* monotonously increase in the future, on the basis of empirical evidence it is only possible to conclude that *in fact* it will experience such an increase. Stated in negative terms, the option of Gold-type universe which Price advertises in his paper, as well as in his recent monograph (Price 1996) is falsified, and should not be taken seriously (except, of course, as a part of the history of cosmology). On the other hand, this issue needs to be investigated in more depth, not only because it might become practical in cosmological future (Ćirković 2002; Bostrom 2003), but also since it highlights a host of epistemological issues specifically pertinent to cosmology to a much larger degree than in most other scientific disciplines.

## 2. Acausal-Anthropic Approach

Once it is understood that the origin of the thermodynamical arrow of time lies (partly or entirely) in the cosmological domain, we need to thread very carefully. Our first motivation in setting up an explanatory project for thermodynamical temporal asymmetry is a full and faithful acceptance of Price's account of the nature of explanatory task (Price 2002a, section 2.2):

> First of all, let's assume that basic explanatory questions are of the form: 'Given that C, why E rather than F?' The thought here is that we never explain things in isolation. We always take something as already given, and seek to explain the target phenomena in the light of that. C represents this background, and E the target phenomenon. (C might comprise accepted laws, as well as 'boundary conditions' being treated as 'given' and unproblematic for the purposes at hand.)

We should, naturally, seek to make C maximally comprehensive. Obviously, our existence as observers is part of the necessary background. Should it not be included in C? Surprisingly enough, Price and most other researchers in the field answer in the negative, and the reason for this is not hard to grasp. In most practical physical applications it is irrelevant, at least in the first approximation (and therefore is usually omitted, although reasons for the omission are often only tacitly implied). Our existence as observers hardly impacts the explanatory project for by far most chemical, geological or solid-state physics phenomena. However, leaving observers out of the picture *in the cosmological context* does not lead to happy consequences, as was first shown by Dicke and Carter in 1960s and 1970s. The outcome of the controversy on the so-called "large-number coincidences" (Ćirković 2003, and references therein) should warn us that for explanation of global properties of our cosmological domain we are not free to leave properties of intelligent observers inhabiting it out of the picture.

With this in mind, we propose a third approach to the explanation of the thermodynamical asymmetry, which could be labeled the Acausal-Anthropic approach. The main physical picture of this approach has been elaborated elsewhere (Ćirković 2003). It is essentialy a Dicke-like approach, applied to the specific problem of the thermodynamical asymmetry and the nature (entropy-wise) of the cosmological initial conditions. To understand what this proposal encompasses, perhaps the best way is to use Price's "parsing" the natural phenomena for the approaches he considers (Price 2002a, §2.3; 2002b, §3.1-3.2). Applied to the Acausal-Anthropic approach, it may look like this:

Symmetric boundary conditions—entropy high in the past

Symmetric default condition—entropy likely to be high (always)

+  Asymmetric observational selection effect

———————————————————————————————

Observed asymmetry

This scheme could be compared with similar parsing schemes for the Causal-General view:

Asymmetric boundary condition—entropy low in the past

+  Asymmetric law-like tendency—entropy constrained to increase

———————————————————————————————

Observed asymmetry

and Acausal-Particular view (Price 2002a):

Asymmetric boundary condition—entropy low in the past

+  Symmetric default condition—entropy likely to be high, *ceteris paribus*

———————————————————————————————

Observed asymmetry.

The basic idea of the Acausal-Anthropic approach is following: having already received from (quantum) cosmology a useful notion of the multiverse, we could as well employ it in order to account for the *prima facie* extremely improbable choice of (local) initial conditions. In other words, we imagine that everything that exists, for which we shall use the term multiverse,[1] represents a "Grand Stage" for unfolding of—among other things—thermodynamical histories of chunks of matter. And the multiverse immediately solve the problem of extraordinarily improbable endpoints (or initial points!) of those chunks we observe in out vicinity. Entropy *in the multiverse* is almost everywhere high at all times ("almost" here means "everywhere minus possible subset of a very small or zero measure"). Our cosmological domain ("the universe") represents a natural fluctuation—presumably of very small or zero measure; but the anthropic selection effect answers the question why do we find ourselves on an upward slope of such a fluctuation. Hence what needs to be explained is not that such fluctuations exist (this is entailed by the Boltzmann measure); nor the fact that the local initial condition is one of an extremely low probability (this results from the multiverse concept); but the fact that we happen to live in such an atypical region of the "grand total" which is almost always at the equilibrium. And this can be explained by showing why the observed entropy gradient is required for our existence as intelligent observers (for further details, see Ćirković 2003).

Obviously, the special place in this approach is reserved for qualificative "observed". Of course, our theories strive for explanation of observed physical world, but it is important to distinguish this truism from the unnecessarily strong realism, which is often used and abused in contemporary science and philosophy. The end results of both this proposal and the other two proposals Price describes are the same: an *observed asymmetry*. However, the attribute seems superfluous in both Acausal-Particular and (especially) Causal-General approaches, since they suffer from unnecessarily strong realism. It has no function at all in either approach. Only in an anthropic approach advocated here, it does receive the proper place in the *core* of the perceived explanandum.

The Anthropic-Acausal approach is indeed what Price (2002b, §3.4) calls the "no-asymmetry" proposal. The label is multiply defficient, however, although the defficiency becomes understandable only when some other manners of defense for the Acausal-Particular approach are considered carefully. Although both "parsing schemes" Price considers (that is, the Causal-General and the Acausal-Particular approaches) contain the same summary line ("observed asymmetry") nowhere has been elaborated on the role (or even meaning) of the attribute *observed*. Is "observed" in this context the same as "objective" or is it not? By equivocating on this, Price attaches a strongly realist (and, indeed, essentialist) character to the two approaches he favors. The picture originating from Boltzmann (and Schuetz) spoil the essentialist fun by taking "observed" in its true and literal meaning (i.e. observed by intelligent observers possessing specific capacities, epistemic and otherwise).

However, we should not accept any possible suggestion that the Acausal-Anthropic view is subsumed by the Acausal-Particular and should not be discussed separately. To perceive the difference between the two approaches, it is enough to ask whether their achieved aims would be any different—*apart* from the fact that they would offer explanations for the thermodynamical asymmetry. Clearly, a full theory developed along Acausal-Anthropic lines would differ from the one developed along Acausal-Particular lines in several respects. Acausal-Anthropic-guided theory would include the postulated multiverse which could be independently checked, not only theoretically, but also empirically (if the domains are not sealed off, but are topologically connected, possibly through wormhole solutions of the full quantum field equations); this would be absent from the Acausal-Particular theory (at least in its "minimal" formulation, which Price seems to favor). Acausal-Anthropic-guided theory, on the other side, would include specific properties of intelligent observers and would have to explain the link of entropy gradient to the functioning of the mind, something which has been entirely beyond the goals and tenets of any Acausal-Particular-guided theory. Thus, we have full right to treat approaches differently (not rejecting their similarity at some points, though). That fact that Price is sometimes unfortunately lackluster with the important methodological issue of demarcation is indicated, parenthetically, in the papers considered here by referring to Ghirard-Rimini-Weber quantum theory as "GRW interpretation of quantum mechanics". Upon no reasonable set of epistemological criteria can GRW be construed as an "interpretation":[2] It has both a different formal structure (a stochastic term in Schrödinger equation) and predicts different observable phenomena (it even has a different set of the constants of nature, compared to the standard view!).

Price's suggestion (2002a, section 3.3) that the anthropic selection effects may be necessary only for explanation of the initial conditions of our domain—and even there it is the option obviously less interesting for him, since it can conveniently be "ignored"—certainly does not do justice to the scope and ingenuity of anthropic reasoning. His preferred option, the existence of law-like reason for low entropy is, in effect, a strange retreat from the *acausal* behavior he correctly emphasizes at the present time. Granting that there may be interesting reasons for exploring this possibility, we are still entitled to ask: why should one, after rejecting law-like behavior of matter at the present epoch, ask for a new law-like behavior in the epochs long gone to explain the same thing, especially if a plausible (and rather more "Boltzmannian") anthropic alternative is at hand? In this respect, Price's version of the Acausal-Particular approach is more complicated and epistemologically demanding than the Acausal-Anthropic alternative advocated here. (Of course, the answer to the question posed above is clear: Price wishes to have a route open for the Gold model and possibity for completely symmetric, low-entropy future. However, as we shall show in the next section, the effort is largely Pigmalionic and—in so far as it leads one to neglect other plausible alternatives—misleading.)

The present approach can be regarded as another instance of the so-called Strong Anthropic Principle (SAP), which has been target of much criticism, since its formulation by Carter (1974). But, contrary to the oft-repeated adage, SAP in the multiverse context is not teleological; quite the contrary, it says only that in a sufficiently broad distribution of a particular cosmological parameter (for instance, the initial entropy-per-baryon) over the entire multiverse, it is unavoidable to have a subset of universes with this parameter arbitrarily close tuned for the subsequent appearance of intelligent observers. This important point has also been argued by Balashov (1990). If there were no multiverse, or in a multiverse of too restricted possibilities, SAP would indeed have been a reformulation of the design argument; fortunately, our explanatory task *presupposes* the right kind of multiverse in which SAP is just a selection effect.

A particular attention should be taken to separate the Acausal-Anthropic approach from the original Boltzmann-Schuetz version. One of the reasons, of course, is that the present approach incorporates much of the development of modern cosmology in both its classical and quantum aspects; those developments were, naturally, almost completely unknown in the time of Boltzmann and Schuetz. The other reason, however, is that some of the objections traditionally raised against the Boltzmann-Schuetz (or "anthropic-fluctuation") picture are invalid as criticisms of the Acausal-Anthropic approach. The most important example here is what in the companion paper (Ćirković 2003) we call "Feynman's Argument" (after Feynman 1965): that we should expect much smaller—in comparison to the size of the "observable universe", for instance— spatial entropy fluctuation to be sufficient for development of the intelligent observers. We consider this counterargument in more detail in the companion paper, but for the moment let us state that there are several ways of defusing it. The most promising one lies in arguing that there are physical reasons—offered by the theory of inflation—why the creation of very small (in cosmological terms) low-entropy domain is impossible or very improbable, and that events taking place only in large universes (like the formation of the large-scale structure and the process galaxy formation) are *causally linked* to our appearance as intelligent observers. As far as astrophysics is concerned, prospects for such a resolution are rather good, since it seems that both nucleosynthesis of elements heavier than lithium and the formation of terrestrial planets are dependent on these "grand stage" processes (Hogan 2000; Oberhummer et al. 2000). In other words, not all types of universes are created out of chaotic inflation (thus we are deviating from Boltzmann's possibility of "all natural events... without restriction"), but they belong predominantly to two types: ones with very large entropy throughout and ones with very low entropy. The exact nature of this distribution (as well as other details of the description of the multiverse itself) will be elucidated upon reaching the much-desired quantum theory of gravity which should give the answer on the "right" form of inflation potential and the proper dynamics of the multiverse (Linde 1990).

## 3. Physical eschatology and implausibility of low-entropy future

A particularly troublesome burden on the rival Acausal-Particular approach—in Price's presentation—are persistent attempts to associate it with the Gold-type cosmological model, which is a recollapsing world with reversal of entropy gradient at the point of maximal expansion (Gold 1962). In this section, we shall try to stop "chasing beauty" and return to sober empirical facts. We have already seen in the course of history of physics how empirical developments have obviated Dirac's "objectivistic" explanation of the classical large-number coincidences, motivated (in the spirit of insisting on elegance of which Dirac was the most vocal promoter in the XX century science) at least partially by philosophical and aesthetic considerations. We should investigate now whether similar fate may befall Gold's time-symmetric universe. On a first glance, it seems that Price highly values the cosmological input in discussions on the origin of the asymmetry puzzle (Price 2002a, §1):

> It turns out that... the Boltzmann statistical considerations in themselves do not give us grounds for confidence that entropy will not decrease in the future. In so far as we presently have such grounds, they rest on what has been learnt quite recently about why entropy was low in the past, and on the implications of this new knowledge concerning the future. (Hence, it turns out, they rest on cosmology.)

So far, so good. However, a deeper look here and elsewhere (e.g. Price 1996, pp. 95-97) shows that Price is extremely reluctant to face such independent evidence. Cosmological revolution taking place in the last couple of years seems to be completely outside of his horizon: the massive evidence for a large positive cosmological constant (e.g. Riess et al. 1998; Perlmutter et al. 1999; Zehavi and Dekel 1999) obliterates prospects for *any* recollapsing universe, and *a fortiori* the prospect for a *very special* case of recollapsing universe, such as Gold's. (This "specialty" does not result from any double standard, but simply from the fact that there are *other* recollapsing world models which are not entropy-reversing; thus, the subset of entropy-reversing models is special within the set of recollapsing models.) Ignoring this development, as well as the entire tradition of observational cosmology in at least 30-odd years of history of attempts to measure the cosmological density fraction $\Omega$, certainly deserves a dictum of Earman (1995, p. 269): "...it needs repeating that philosophy of science quickly becomes sterile when it loses contact with what is going on in science."

One of the small number of pieces of empirical data which should certainly be taken into consideration in any discussion of the asymmetry of time is the fact that today we have overwhelming evidence that the universe will not recollapse; apart from that, there are also theoretical reasons why the closed universe is not an appealing option any more (as it may have been in the time of Einstein or Tolman).[3] The list of references containing the empirical evidence for

(1) $\Omega_{matter} < 1$, and

(2) $\Omega_{\Lambda} > 0$,

is certainly far beyond the scope of the present manuscript; for a bit of flavor, one may mention works of Bartlett and Blanchard (1996) on the cosmic virial theorem; Fan, Bahcall and Cen (1997) on mass indicators in galaxy clusters; Bertschinger (1998 and references therein) on the large-scale velocity maps—all strongly supporting (1). Gravitational lensing, one of the most powerful astrophysical tools ever devised, strongly supports (1), as testified—among other studies in this very active field—by Kochanek (1995). The case for a topologically open universe (which implies that it is also ever-expanding, although the converse is not true, provided that the sign of the cosmological constant is positive, as usually assumed, since even the topologically closed universes can also expand forever) has been reviewed by Coles and Ellis (1994) and found to be quite solid, even before the findings of cosmological supernovae projects (Riess et al. 1998). Even without these empirical results, the onus of proof should lie with proponents of ¬(1), since nobody has ever observed anything even remotely similar to the quantity of matter necessary for *any* type of recollapsing universe. For example, all shining stars, detectable interstellar gas and dust comprise about one thousandth of the mass density which is the necessary condition for recollapse. While everybody is free, of course, to postulate any sort of super-duper hidden dark matter with sufficiently homogeneous distribution, it is rather obvious that such theories have to be deeply contrived in order to account for the existing empirical database (and that the burden of proof lies entirely upon proponents of such theories).

Apart from that, observational discovery of the large positive cosmological constant—point (2) and several theoretical considerations preceding the empirical discoveries—offers exactly what Price mentions in the very next paragraph: a plausible reason for rejecting recollapsing universes *in general* (Riess et al. 1998; Perlmutter et al. 1998, 1999). Now, one may argue that an argument is insufficient unless the complete quantum field theory (including gravity) is attained, wherein the vacuum energy density would be calculable "from the first principles". However, we find the remark to be rather peripheral; the theoretical fact that the cosmological constant is a *natural* product of field theories and the observational fact that the cosmological constant seems to govern the universal expansion are sufficient enough to doubt whether any universe could be closed (except by a highly improbable and unappealing numerical conspiracy). Apart from observations of cosmological supernovae which brought forth this true revolution in cosmological thinking, there are other pieces of empirical data pointing in the same direction (and giving surprisingly close numerical best-fits), the most profound of them being the observations of the small-scale anisotropies in the cosmic microwave background radiation (Lineweaver 1998). Anthropic reasons for predicting a positive cosmological constant have been put forward by Efstathiou (1995), and the entire richness of theoretical motivation for $\Lambda$ reviewed in detail by Martel, Shapiro and Weinberg (1998). Thus, neither observations nor cosmological theory favor *any* type of recollapsing universe—without mention of the Gold universe, which is a very *special* type of the recollapsing universe. In light of all this evidence, it is somewhat surprising why the closed universe is still popular, in particular outside of cosmological circles; Price offers just a recent and highly sophisticated example of that tendency. On the contrary, observational astronomers have persistently warned that our universe is in all probability an open, eternally expanding one; there's practically not a single case in the entire history of observational cosmology, seven decades from Hubble to this day, that an observational astronomer suggested a closed, recollapsing universe (Brawer and Lightman 1990). Of course, issues like this cannot be settled by a majority vote, but it is indicative of the overall state of affairs; and the majority vote of experts in an empirical issue is certainly as good a guideline as any aesthetic or symmetry postulate or one's perception how the integral "Archimedean" worldview should look like. This is especially valid if one confidently quotes "cosmological evidence" whenever it is convenient from the point of view of one's own approach. It may be speculated that there are aesthetical, sociological or even religious factors involved in the attraction to the closed universe case[4] (paradigmatic instance being Tipler's Omega-point theory; see Tipler 1994), but weighting of these extra-cosmological arguments is beyond the scope of the present discussion.

It should be stressed that relatively new findings concerning the large positive cosmological constant point essentially in the same direction as the previous cosmological lore, in spite of the picture that Price and some other authors assume. The grounds for reintroducing the cosmological constant have existed for decades, as testified by the thoughtful review of Carroll, Press and Turner (1995) predating the discoveries of Perlmutter et al. and the cosmic microwave experiments. A particularly strong case for a large positive $\Lambda$ from the data on astrophysical ages has been made by Roos and Harun-or-Rashid (1998) in a paper appearing slightly before the main supernovae results. All these (and others, the full bibliography might as well be longer than this manuscript!) empirical data come on the top of the following strong theoretical reason: the positive cosmological constant is an almost mandatory product of the quantum field theories which are considered so highly respectable in other cosmological application (for instance, in explaining the early inflation). The fact that this family of models was historically investigated to a lesser degree than Einstein-de Sitter $\Omega_{matter} = 1$ model testifies, perhaps, more on the simplicity and lack of diligence of theoretical cosmologists, as well as other elements of the sociology of science, than on the nature of the universe. Ever-expanding universe has never been "news".

A word should be said here about the relationship of epistemology and physics in cosmological matters. In his book, Price subtly shifts grounds claiming that (p. 95):

> The main issue is whether the Gold view makes sense in a recollapsing universe, not whether our universe happens to be a recollapsing universe. Of course, if we could show that a recollapsing universe is impossible, given the laws of physics as we know them, the situation would be rather different: we would have shown that the original puzzle concerns a case that physics allows us to ignore.

However, one may find a host of examples in Price's book (as well as the two discussed articles) that indicate different attitude toward interpretation of physical results in the light of an atemporal worldview. For instance, when discussing the macroscopic notion of entropy increase in Chapter 2, Price (1996) takes for granted that it is Boltzmann kinetic theory or statistical mechanics in general that is governing the real world. Accordingly, there were no conditionals of the form "We are interested whether the world in which macrosystems are composed from a huge number of identical simple subsystems, etc. is a coherent model, not whether the real world can be described in such a way, and the issue whether actual entropy increases or not is rather peripheral", etc. The persistent usage of the location "the world" also suggests a rather realist view on his part. In the light of cosmological evidence discussed above, the question "Will entropy be high or low near the Big Crunch?" is similar to the question "Is there more phlogiston in acetylene compared to cerosene, or vice versa?" Neither question is *logically* senseless; there may be a context in which both questions could be coherently answered. Nevertheless, as questions about actual *physical reality* they are meaningless, since the progress of empirical science has obviated the entities involved (Big Crunch and phlogiston, respectively).[5]

On the positive side, physical eschatology enabled us to get some hold on the issue of future entropy production. Thus, for example, in one of the first studies of cosmological future (entitled "Entropy in an Expanding Universe"), Frautschi (1982) wrote:

It is apparent... that the entropy in a causal region falls steadily further behind max S during most of the cosmic history. $S/S_{max}$ does increase temporarily during the period of stellar and galactic black hole formation. Life as we know it develops during the same period, utilizing the much smaller, but conveniently arranged entropy generation on a planet or planets situated near nucleosynthesizing stars. ...the expanding universe does "die" in the sense that the entropy in a comoving volume asymptotically approaches a constant limit.

In the similar vein, Adams and Laughlin (1997), in the comprehensive study of almost all aspects of physical eschatology, notice that:

Thus far in this paper, we have shown that entropy can be generated (and hence work can be done) up to cosmological decades $\eta \sim 100$. [Cosmological decades are defined as epochs of time $t = 10^{\eta}$ years] ...The formation of larger and larger black holes, can continue as long as the universe remains spatially flat and the density perturbations that enter the horizon are not overly large. The inflationary universe scenario provides a mechanism to achieve this state of affairs, at least up to some future epoch... Thus the nature of the universe in the far future $\eta \gg 100$ may be determined by the physics of the early universe (in particular, inflation) at the cosmological decade $\eta \sim -45$.

But it is not necessary to involve the wealth of technical details of these and related studies here; it is enough to point out that there already exists a sizeable volume of scientific literature on the entropy production in the cosmological future—literature Price conveniently chooses to ignore.

Examples of neglect of these considerations abound. For instance, in section 9 of Price (2002b) we read that:

... it hasn't been established that entropy will continue to increase. We saw that what has been added since Burbury's time is the cosmological case for regarding this as an open question. Burbury was rightly suspicious of the crucial assumption (the assumption that he himself had first identified) of the leading argument for thinking that entropy will continue to increase. The cosmological connection supplies an independent reason for holding open the possibility that it will not do so.

The first trick here is equivocating on "established". If "established" means "established with 100% certainty," or "without any doubt," or "mathematically proven" then the first sentence is correct. Only a full-blown and operational "Theory of Everything" could establish this (or its opposite). However, if we—more appropriately for a physical science which is seldom so rigid—take this term as, say, "indicated by all empirical evidence, without a likely alternative", then Price's statement is incorrect and misleading. However, the cover-up of the real issue continues when the author claims that what Burbury debated with is "the leading argument" for the prediction of entropy increase. "Leading arguments" for this stem from modern astrophysics, about which Burbury has been, of course, completely ignorant. The "leading argument" might be, for instance, the fact that $\Omega_{matter} < 1$.[6] Finally, from the remote possibility that modern astrophysics got it all wrong (an option which is always in the cards in science!)[7], Price draws a conclusion that it is cosmology that gives the warrant to a strange idea of future entropy decrease. One cannot escape fascination with this casual juggling with cosmological evidence. However, one also cannot escape the application of Earman's sad conclusion, quoted above.

Another example of this selective (ab)use of cosmological evidence is seen in §8.3 of the same paper (Price 2002b):

We simply *don't* [Price's emphasis] have the particularly strong reasons for believing that entropy will continue to increase, at least in the distant future. However, although the sentiment is Burbury's, the present justification for it depends on considerations of which Burbury could not have been aware. It depends on our current understanding of the nature of the past low entropy boundary condition, and hence on modern cosmology... Given this reason to think that the low entropy past has cosmological origins, we also have some basis for thinking
that the issue of future entropy also turns on cosmological issues.

This (and subsequent paragraph) is doubly misleading: first in blatantly selective
approach to the bulk of the modern cosmological evidence (which is allegedly held in
high esteem); and secondly, in claiming that there is, historically, any substantial
*cosmological* issue motivating the option of low-entropy future conditions. Gold's own
exposition of the idea was partly—on which he was rather explicit—based on his
personal aesthetical and philosophical preferences, and partly on the idea that the
Hubble expansion literally increases entropy through increase in the accessible region of
phase space—the idea that has been *proven wrong* (e.g. Layzer 1976). Gold's idea was
also encouraged by the erroneous belief, widely held at the time, that there were no
fundamental time-asymmetric processes. And that is about everything we have for the
low-entropy future: some personal tastes, some wishful thinking, and some fallacious
cosmological ideas. And, of course, aesthetic tastes could be exactly the opposite with
no lesser scientists (see the endnote 4, for instance).

This certainly has grave consequences for the other parts of Price's account. In
discussing the options for explaining the atypical initial conditions of our cosmological
domain, Price, as we have already noted, prefers some sort of law-like correlations, so
that these conditions are no longer exceptional. However, as we have discussed in the
previous section, this violates the commitment of any acausalist to avoid new laws of
nature as long as possible. In answering this objection, Price writes (2002a, sec. 3.3):

> Wouldn't it amount to a revision of our background probability
> measure on the space of possible histories for the universe, so that this
> measure would no longer be time-symmetric?... No, for two reasons.
> First, it need not make the global probability measure time-
> asymmetric. It might turn out to have the consequence that entropy
> would decrease towards a Big Crunch, as well as towards the Big
> Bang. In this case, the asymmetry of the measure in our vicinity would
> be balanced by a reverse asymmetry elsewhere.

Obviously, this is again a counsel of despair. If we accept the cosmological evidence at its face value and reject the reality of Big Crunch, this tactics fails to convince us that the global probability measure would retain temporal symmetry on Price's approach—that is, unless we are willing to extend the measure to other cosmological domains and accept the anthropic selection mechanism. Is that really so high a price that we should rather consider an *ad hoc* undiscovered law of nature instead?

Of course, it should be emphasized here that the Acausal-Particular view promoted by Price should not be necessarily associated with any form of the entropy-decreasing future models. What merits the Acausal-Particular approach may have or not have to the explanation of the thermodynamical asymmetry, ought to be judged on its features alone, and not on the rather casual associations with a dubious cosmological view such as Price's. The best thing one may say is that—in contradistinction to the Causal-General approach—this approach *allows* for a low-entropy future. Any stronger statement (for instance, that the Acausal-Particular approach somehow *implies* low-entropy future boundary) constitutes a misprepresentation and can be only harmful for this, otherwise quite legitimate, approach.

Let us note finally that on Acausal-Anthropic view advocated in §2, the Second Law needs not *necessarily* hold true in the cosmological future (this point is common with the Acausal-Particular approach of Price). This is hardly necessary to explicate if one accepts the validity of the Boltzmann measure on its face value. However, in our view, the nature of future boundary conditions is an empirical issue, and on physical eschatology to decide. No oracle (or "Grand picture" or crystal ball or aesthetically appealing symmetry) can provide us with an answer in advance of observations and modeling. For the moment, prospects for any future entropy-decreasing behavior are extremely dark (arguably, they were never really bright at all). We have all reasons to believe that the universe will expand forever, in an accelerating pace, while the matter will gradually either decay (through GUT proton decay) or annihilate (as electron-positron pairs) or be swallowed by black holes, only to be unimaginably slowly radiated away through the Hawking evaporation mechanism. In addition, there may exist a connection to the properties and capacities of intelligent observers, which could be different in the future from the ones in the past, as was suggested by Dyson (1979) in his pioneering study in physical eschatology (but see Oppy 2001). The situation may be different as well if the global *topology* is not trivial (Ćirković and Bostrom 2000).

## 4. Conclusions

We conclude that a third approach, which we have labelled Acausal-Anthropic approach may be able to account for the perceived thermodynamical asymmetry at least in the same degree as the two approaches explicated by Price (2002a,b). An enormous additional benefit comes, of course, from being able to account for the other anthropic coincidences (or 'fine-tunings') within the same conceptual framework received from the rapidly developing ideas of quantum cosmology. Further advantages include a connection with the latest thinking in quantum cosmology (incorporating the idea of the multiverse), dropping off the *ceteris paribus* clause in the specification of the default thermodynamical condition, better accounting for thermodynamical counterfactuals and obviation of a necessity to double-deal with cosmological data in the way Price does in pleading for the Gold-type universe. At the same time, many of the virtues of the Acausal-Particular approach are retained in the Acausal-Anthropic picture. While the work of Price represents a significant step forward, especially as far as criticism of the Causal-General type of explanation goes, he has not entirely fairly portrayed the situation as far as the search for explanation of thermodynamical asymmetry is concerned.

Future behavior of the universe is rapidly becoming a recognized and legitimate target for "everyday" scientific work, and less and less an arena for wild speculation and "grand principles" (Oppy 2001). The nascent discipline of physical eschatology has already reached many interesting results, and it is highly misleading to present contemporary astrophysicists as ignorant about the subject as their colleagues in the time of Boltzmann or Haldane or Gold. While healthy critical attitude is more than welcome, simply ignoring recent revolutionary developments in cosmology (and in particular the confirmation of existence of a large vacuum energy density) is hardly a praiseworthy standpoint. According to an unequivocal conclusion drawn from these empirical developments, (asymptotic) final state of our cosmological domain (or "universe") will be one of extremely dilluted matter and extremely high entropy, thus obviating the sterile possibility of future anti-thermodynamical behavior (at least as long as one keeps intentional actions of advanced intelligent communities out of the picture).

**Acknowledgements.** We acknowledge three anonymous referees for helping significantly improve the previous version of the manuscript.

---

[1] Not to be confused with the multiverse of quantum mechanics ("Everett's multiverse" = the totality of wavefunction branches). Here we refer to the multiverse of quantum cosmology ("Linde's multiverse" = set of different cosmological domains, possibly causally and/or topologically disconnected from ours).

[2] Neither should Everett's quantum theory (since it also predicts novel phenomena), although it is less clear-cut case; for two differing views on this issue see Deutsch (1996) and Tegmark (1998a).

[3] Although we cannot deal much with these theoretical issues here, we can direct the reader to the study of Barrow and Dabrowski (1995), as well as earlier discussions of Landsberg and Park and others discussing oscillating universes. In addition, closed universes face conceptual as well as observational problems connected with the turn-around images of light sources, gravitational megalensing, and some other issues, which they could escape only if they are closed by a very small margin, $\Omega_{total} = 1 + \varepsilon$. Here, as well as in the entire text we use the common cosmological notation of dimensionless density fraction of energy component $i$ as $\Omega_i \equiv \rho_i/\rho_0$, where $\rho_0$ is the critical density for the universe to recollapse. If the universe contains only matter fields, it will expand forever for $\Omega_{matter} \leq 1$ and recollapse to the big crunch for $\Omega_{matter} > 1$. According to the modern view, the total cosmological density $\Omega_{total} = \Omega_{matter} + \Omega_\Lambda$, where $\Omega_\Lambda$ is the vacuum energy density commonly known as the cosmological constant (or, more generally, "dark energy"). As elaborated in a beautiful essay of Krauss and Turner (1999) cosmological constant introduces a degeneracy into cosmological future, which indicates that even topologically closed $\Omega_{total} > 1$ can expand forever in the presence of positive cosmological constant.

[4] Which Dyson, parenthetically, labeled "claustrophobic model" (Dyson 1979).

[5] Other investigators, even those favorably disposed toward time-symmetric models, admit that much. For instance, Kent (1998) cautiously remarks that "there are other reasons to approach... time-neutral cosmologies with skepticism... The evidence appears to lean against a closed universe, and a closed universe is natural cosmological setting for time-neutral models."

[6] The same applies to the remark in footnote 21 of Price (2002a), where we read that "the problem is that we are not justified in postulating such a law, unless we have independent reason for excluding the possibility of a low entropy future." Well, $\Omega_{matter} < 1$ seems independent enough from the sort of the statistical and thermodynamical discussion listed in Price's discourse.

[7] But the option that astrophysics got it all wrong should only be weighted against the option that we got the fundamental dynamics wrong, and that it is not completely time-symmetric after all (which is recognized within the Causal-General approach). The latter case could be even more promising if one insists on rare, but empirically established, phenomena like the minuscule CP violation.